## An Operational Approach For Wimax At Ultra High Bandwidth With Spectrum 60 Ghz

S. SOOMRO, A. A. KHAN*, A. G. MEMON**, A. IFTIKHAR, M.R MAREE**,

Institute of Business Management, Karachi.
Email: safeeullah.soomro@biztekian.com  ghafoor@usindh.edu.pk

**Abstract**: WiMax is a promising network of today industry. It provides P2P and P2MP point to multipoint broadband services up to thirty miles. Its operational frequency range is 10 GHz to 60 GHz. It provides data rate of 75Mbps per channel; with an end-to-end encryption called CCMP (Counter Mode with Cipher Block Chaining Message Authentication Code Protocol). CCMP is an Advanced Encryption Standard AES based encryption method, which delivers secure communication. Telecom industry seeks secure, cheaper, wireless metro area network, that full fill the today internet demand in most efficient way. Our research explores the new dimension of WiMax with HEMT High Electron Mobility Transistor using un-licensed Band. The Aim of this paper is to simulate 60GHz unlicensed WiMax band in Matlab. This research explores millimeter waves, related electronics, mathematics of WiMax and simulated graph comparison. It determines available capacity verses coverage area and transmission bit error probability. Further it highlights the ideal modulation condition in different terrains. Research proofs that WiMax will be the future promising wireless network.

## 1. INTRODUCTION

Wireless networks are economical and portable. It market ratio is 1:3. Wireless networks are famous among consumers due to secure, reliable, robust and portable connections. WiMax network is one of the famous networks among users; it is provides excellent multimedia services. WiMax is a (Worldwide Interoperability for Microwave Access) which is IEEE 802.16 standard supports licensed and unlicensed bands ranging from 10 to 110 GHz. It is reliable due to availability of signal up to 99.999%. This technology can further be improved by installing HEMT High Electron Mobility Transistor in its antenna. HEMT technology proves that it has very high gain, low noise figure in millimeter wave frequency range. W-band low noise amplifier is a key component of next generation satellite communication systems, wireless LAN, and radio telescope receivers. Some W-band low noise amplifiers developed on GaAs based HEMT. These low noise amplifiers demonstrated better gain and noise figure performances. This technology works on unlicensed band and provides metro area coverage (Arafat, and Dimyati, 2010, Jeffrey , *et al.,* 2007).

It can be a mesh network. WiMax is the combination of the following core technologies such as Orthogonal Frequency Division Multiplexing OFDM, Low Density Parity Check LDPC, Multiple Input and Multiple Output MIMO/Smart Antenna and Software Defined Ratios SDR Kyung-Ho Kim, (2006). WiMax can deliver high speed data rate upto 10 Gigabyte/s over five Kilometers in LoS Line of Sight network. For non-line of sight coverage it prefers Orthogonal frequency division multiple access OFDMA (Arafat, and Dimyati, 2010) High Electron Mobility Transistor (HEMT) technology is behind this efficient network. It has four and six-stage 60 GHz low noise High Electron Mobility Transistor (HEMT) amplifier, which amplifies and generates millimeter waves. The fourth-stage HEMT version is designed for beamed signal scattering with less noise, with an associated gain of 15.1 dB at 60.2 GHz frequency. Likely, six stage HEMT device provides 4.9 dB of associated gain, with an increase of 21.7 dB at 61.5 GHz frequency. This study proofs that 0.25µmeter Pseudo-Morphic High Electron Mobility Transistor (PM-HEMT) technology is an appropriate choice. It is cost-effective,V-band LNA (Low Noise Amplifier). These LNAs are practically installed in satellites for inter-satellite communication. This research combines WiMax protocol with HEMT technology hardware that covers 50 kilometer area. This combination can be a mesh network based on predefined set of instruction (Jeffrey , *et al.,* 2007, Santhi, Senthil Kumaran (2006) WiMax 60GHz technology uses 0.25 µmeter PM HEMT version Low Noise Amplifier for P2P communication; this is LOS based technology

++ Corresponding author: email: adnan.alam, asim.iftikhar}@iobm.edu.pk
*Institute of Business and Technology, BIZTEK, Karachi, Pakistan
**Institute of Mathematics Computer Science, University of Sindh, Jamshoro, Pakistan

and covers an area of 3 Km with 1GByte data rate. HEMT is millimeter wave technology possess less signal interference, mitigation, advanced encryption, traffic prioritization and provides quality service at higher frequency band. In contrary 60GHz signals absorb in oxygen $O_2$ molecules, works on line of sight only and its coverage area is linked with line of sight. There are four modulation schemes namely: BPSK, QPSK, 16QAM and 64 QAM. Millimeter Antenna is frequency independent and works on geometries of specified angles. It is practically independent of frequency for all frequencies above certain value and predefined impedance. The general formula for their shape is $r = e^{a(\varphi+\varphi_0)} f(\theta)$

Whereas $r, \theta, \phi$ are spherical coordinates, $a, \phi_0$ are predefined constants and $f(\theta)$ is the function of $\theta$. Let "a" is a positive number, '$\phi$' phi ranges from $-\infty$ to $\infty$ which determines the low frequency limit. In these antennas, frequency variation is related with the rotation of the antenna.

## 2. METHODOLOGY

Higher frequencies possess relatively shorter wave lengths. The wavelength of 100 GHz signal is 0.33 cm; in contrast a 100 MHz signal wave has three meters length. Microwave system uses point to point communication, which is mainly between transmitter and receiver. Free space path loss is one of the important terms used to explain microwave transmission. Straight lines emerging from electromagnetic waves go through a vacuum with no reflection or absorption of energy. Effective isotropic radiated power (EIRP) is the related with focused signal radiation, it might be a beam, dish, radio telescope etc. Mathematically.

(EIRP) = (Transmission Power) losses - Cable Losses + (Transmission Antenna) Gain.
(Receiver Signal) Losses = [EIRP] – [FSL] + [RX Antenna Gain] – [Coaxial Cable Loss]

Free space path loss is a term related with antenna gain, distance and frequency. Mathematical equation for free space path is

$$(L_P) = \left(\frac{4\Pi D}{\lambda}\right)^2 \text{ Whereas } \lambda = \frac{C}{f} \quad (1)$$

Parameters for aforementioned formula are as follows distance will be 3Km, frequency is 60 GHz. Hence the calculated path loss will be 137.50. This path loss is a fixed loss, which remains constant over time. Fading is signal loses its strength as much as 30dB with respect to time. Fade margin determines the characteristics of signal strength propagation; it also defines multi path propagation and environmental constraints. Barnett-Vignant reliability equation for specified annual system availability for an unprotected, non-diversity system yields:

$F_m = 30 \log D + 10 \log(6ABf) - 10 \log(1-R) - 70$
(Multi-path effect)(Terrain sensitivity)(Reliability Obj)(K)

Karachi $F_m = 30\{\log(30)\} + 10\{\log(6)(4)(0.5)(60)\} - 10\{\log(0.0001)\} - 70$
$F_m = (44.31) + (28.57) - (-50) - (70)$
$F_m = 53 \text{dB}$

Millimeter antenna is an electronic device which transmits electromagnetic signal from one point to another point/multipoint. The equation for free space propagation between two antennas is given by the Friis transmission equation is as follows:

$\left(\frac{P_R}{P_T}\right) = \left(\frac{G_T G_R \lambda^2}{(4\pi R)^2}\right) g_T(\theta, \emptyset) g_R(\theta', \emptyset')$ whereas $P_T$ and $P_R$ are received and transmitted power, $G_T$ and $G_T$ is the transmitted and received antenna gain, $\lambda$ is the wave length of the electromagnetic waves, R is the distance between two antennas, $g_T(\theta, \emptyset) g_R(\theta', \emptyset')$ the relative directive gains at spherical angles $(\theta, \varphi)$ measured from the pointing direction of each antenna with a convenient reference direction for $\varphi$. The power received during transmission is mathematically expressed in decibels and their relationships are as follows[4][5].

$: P_R = P_T \left(\frac{G_T G_R \lambda^2}{(4\pi R)^2}\right) g_T(\theta, \emptyset) g_R(\theta', \emptyset')$

$P_R = 10\log_{10}(P_R) = 10\log_{10}\left(\frac{P_T G_T g_T G_R g_R \lambda^2}{(4\pi R)^2}\right)$

$P_R = P_T + G_T + g_T + G_R + g_R + 20\log_{10}(\lambda) - 20\log_{10}(R) - 22$

$L = P_T - P_R = -(G_T + g_T + G_R + g_R + 20\log_{10}(\lambda) - 20\log_{10}(R) - 22)$

$L_B = -(20\log_{10}(\lambda) - 20\log_{10}(R) - 22)$

Whereas (L) is transmission loss, (LB) is the basic transmission loss, (P) is power in (dB) decibels, (G) is antenna gain in dB and relative isotropic antenna in (dBi) which is depends upon range and wavelength. The basic transmission loss in free space between two isotropic antennas is as follows.

$P_R = P_T + G_T + G_R - 20\log_{10}(f) - 20\log_{10}(R) - 62.4$

The following equation depicts the efficiency of directive gain of an antenna 'eD' and its relationship is: $eD(\theta, \emptyset) = G_g(\theta, \emptyset)$

The ideal spreading of EM waves determines by three methods one is propagating energy of wave, the other one is far-field directive gain of each antenna, and third one is efficiency of an antenna. The following equation is used for one way communication between two antennas.

$\left(\frac{P_R}{P_T}\right) = \left(\frac{G_T G_R \lambda^2}{(4\pi R)^2}\right) g_T(\theta, \emptyset) g_R(\theta', \emptyset') m. \, 10^{-(0.1(A_T + A_R + \int_0^R \alpha dr))}$

Antenna signals loses its strength due to multiple environmental factors which is generally represented by 'AR', 'AT' its unit in (dB), "m" is a unit that elaborates the signal reduction due to a polarization mismatch between the receiving antenna and sending antenna, "α" is the exact attenuation (dB/km for r in km) due to atmospheric processes along the propagation path.

$$E = E_0 \frac{e^{-j(kR-\omega t)}}{(4\pi R)^2}$$

$$E = E_1 + E_2 = \frac{e^{j\omega t}}{4\pi}\left(E_{01}\frac{e^{-j(kR_1)}}{R_1} + E_{02}\frac{e^{-j(kR_2)}}{R_2}\right)$$

$$P_R = \frac{\lambda^2 G_R}{(4\pi)^2} = \left(\sqrt{P_{T1}G_{T1}g_{T1}g_{R1}}\frac{e^{-j(kR_1)}}{R_1} + \sqrt{P_{T2}G_{T2}g_{T2}g_{R2}}\frac{e^{-j(kR_2)}}{R_2}\right)$$

$$\frac{P_R}{P_T} = \left(\frac{\lambda}{4\pi}\right)^2 G_T G_R \left(\frac{g_{T1}g_{R1}}{R_1^2} + \frac{g_{T2}g_{R2}}{R_2^2}\right) + 2\frac{\sqrt{g_{T1}g_{R1}g_{T2}g_{R2}}}{R_1 R_2}\cos(k(R_2 - R_1))$$

Results can be determined by aforementioned relationship. Software uses this relationship and generates graphs. Results are as follows.
Transmission signal (Tx) 63 GHz and Receiver signal (Rx) is 57 GHz.

a. The 'Rx':

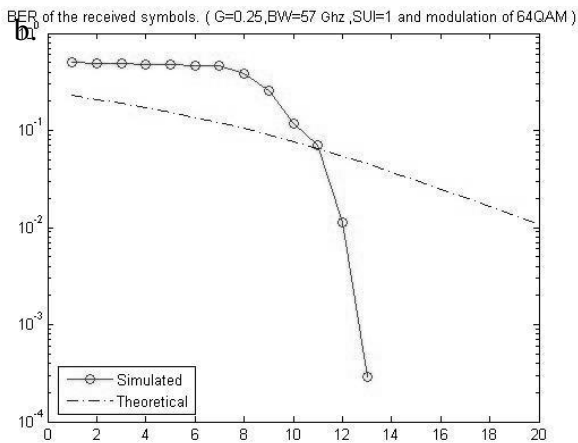

b.

**Fig. 1** This graph depicts two types of 64QAM modulations one is Theoreticalvaluesand other one is simulated valueson57GHz at receiver end. It shows the theoretical value covers 20 meters as compared to simulated value will cover only 13 meters.

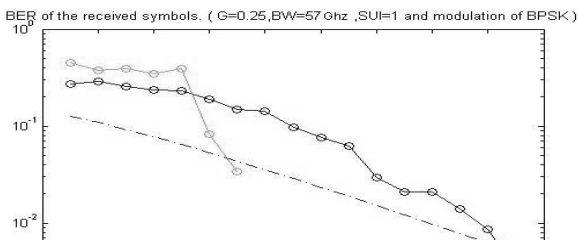

**Fig. 2** depicts QPSK 57GHz BER at receiver end modulations at 57 GHz, its first graph is used without encoded data that covers almost 17 m, second type is encoded data modulation which covers 7 meters and the theoretical data covers beyond 18m.

An aforementioned graph compares simulated and theoretical values of BER of BPSK mode of 57GHz transmission band. The following graph compare BER with QPSK and its related graph is as follows.

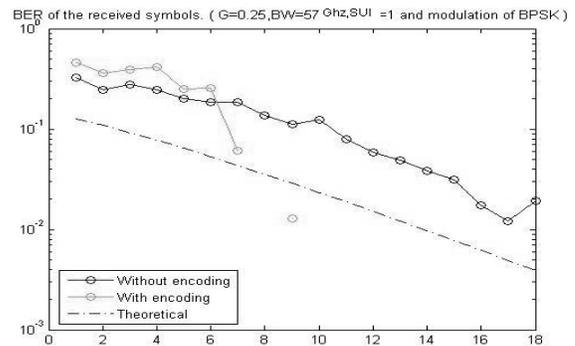

**Fig. 3.** The following graph depicts16-QAM modulation at 57GHz.It shows three types of data encoded, without encoded and theoretical. The encoded data covers above 18m whereas encoded data covers only 7m And theoretical data covers above 18m.

The forth coming graph compares BER theoretical values with simulated values. This graph presents coded data and normal data packets in 16-QAM, 64-QAM communications.

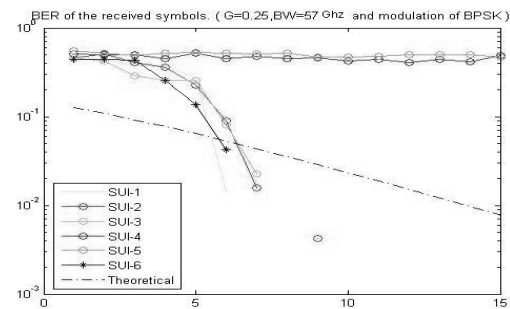

**Fig. 4:** This graph depicts 64-QAM modulation at 57GHz at re receiving end. It shows SUI 1 to 6 and theoretical value in which SUI5 and SUI6 covers above 15 m.

There are two ways of communication 'Tx' and 'Rx'. There communication frequencies are different.

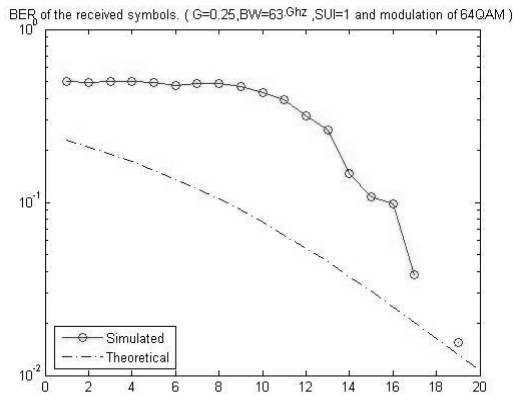

**Fig. 5. This graph shows 63GHz BPSK modulation at Receiving end. It shows simulated value is just Like theoretical value and it covers above 20m.**

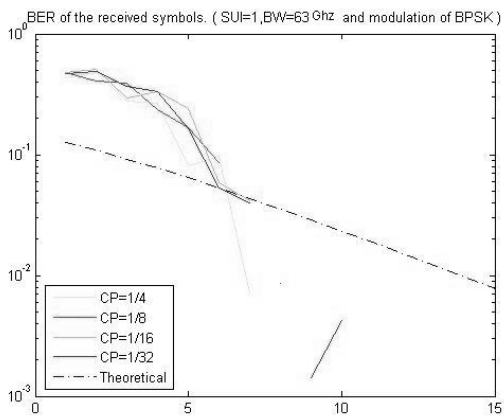

**Fig.6. This graph shows 63GHz QPSK modulation at transmission end. Its CP ¼ covers 8m and its theoretical value covers above 15 m.**

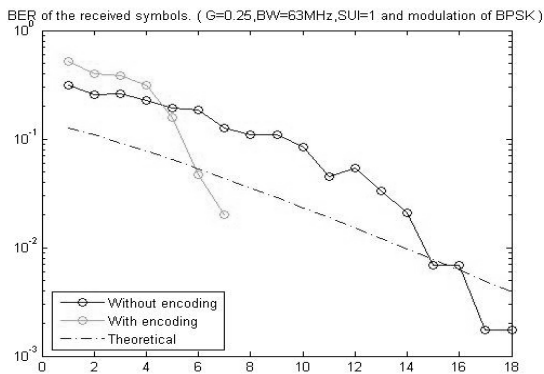

**Fig.7. This graph shows 63GHz transmission At 16QAM. There are three types of modulation encoded, without encoded and theoretical. Encoded data covers above 7m whereas without Encoded covers 16m and theoretical data covers 16m.**

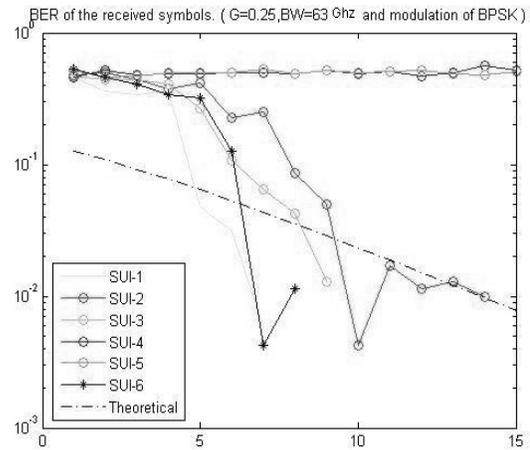

**Fig. 8. This graph shows 63GHz modulations at 64QAM transmission at transmitting end in which SUI2, SUI5 covers area better than theoretical data.**

This study suggests that 64QAM bit error rate (BER) results are better than others. The graphs determines that 64 QAM WiMax modulation data rate is higher than rest of three. It generates less BER and proves its efficiency. This research also highlighted the new dimension in WiMax and urges to use HEMT technology. It seems to be more practical when we combined transmission and receiving gain. In multipath channel same data rate is an issue, it can resolved if fade margin 'Fm' is controlled. The bit error probability for (BFSK) is as follows.

$$P_b = \frac{1+K}{2+2K+\overline{\gamma}_b} \exp\left(-\frac{K\overline{\gamma}_b}{2+2K+\overline{\gamma}_b}\right).$$

Whereas 'K' are Ricean K-factor its range 0→∞ and $\overline{\gamma}_b$ is average (energy/bit) :( noise).

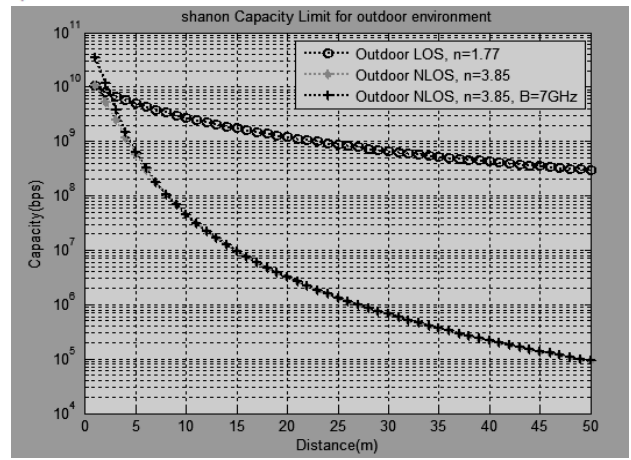

**Fig. 9. Capacity vs Distance This graph shows Shannon capacity vs distance in terms of line of sight and Non line of sight communications, as you can see that line of sight outdoor coverage is good as compared to NLOS.**

Study proves that it is prohibitively impractical to achieve bit error probability $10_E12$ in Ricean and Rayleigh fading channels. Other techniques can reduce fade margin namely coded systems, diversified systems; higher antenna gain system. To create diverse network maximum ratio can be used in combination of flat Rayleigh fading channel, the relationship is expressed as.

$$p_b = \left[\frac{1}{2}(1-\mu)\right]^L \sum_{k=0}^{L-1} \binom{L-1+k}{k} \left[\frac{1}{2}(1-\mu)\right]^k$$

Whereas 'L' means diverse independent available channel.

$$\mu = \frac{\bar{y}_c}{\bar{y}_c + 2}$$

Where '$\gamma c$' is the average Signal to Noise Ratio per channel. High definition video data rate communication at 60 GHz frequency requires blockage free link margin. Further receiver antenna gain values for aforementioned equation will be, 38,50,14 dBi respectively.

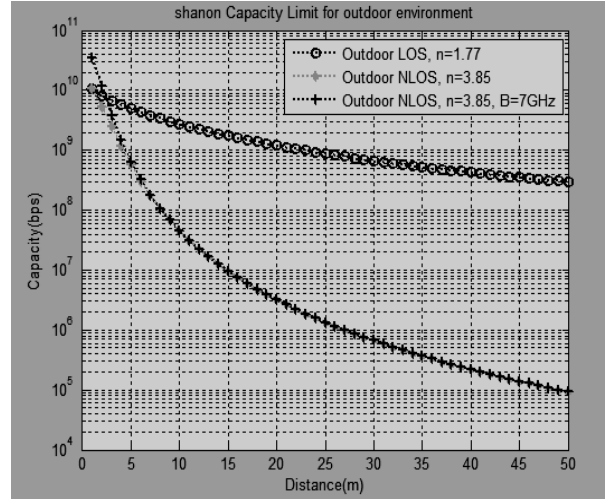

**Fig. 11. Coverge vs Capacity**

**Table 1: Modulation Comparison specifications**

| S.No | Modulation | Channel Type | Band width | Data Rate | Cyclic Prefix 'G' | Simulation Time |
|---|---|---|---|---|---|---|
| 1 | BPSK | P2P | 57 GHz | 160 Bits/sec | 1/4 | 240 |
| 2 | QPSK | P2P | 57 GHz | 160 Bits/sec | 1/4 | 11.431 |
| 3 | 16QAM | P2P | 57 GHz | 160 Bits/sec | 1/4 | 14.876 |
| 4 | 64QAM | P2P | 57 GHz | 160 Bits/sec | 1/4 | 1.1683 |
| 5 | BPSK | P2P | 63 GHz | 160 Bits/sec | 1/4 | 3.5859 |
| 6 | QPSK | P2P | 63 GHz | 160 Bits/sec | 1/4 | 40.198 |
| 7 | 16QAM | P2P | 63 GHz | 160 Bits/sec | 1/4 | 14.729 |
| 8 | 64QAM | P2P | 63 GHz | 160 Bits/sec | 1/4 | 61.764 |

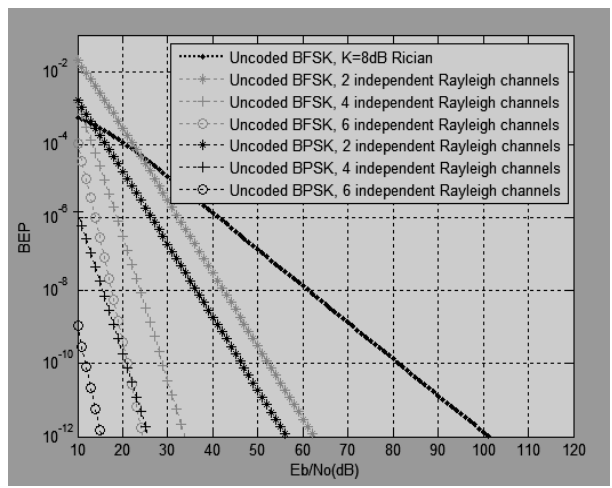

**Fig. 10. Probability vs dB**

### 3. CONCLUSIONS

This study explores WiMax modulation at 60GHz frequency band. It generates comparative graphs among BPSK, QPSK, 16QAM and 64QAM. Further it highlighted HEMT technology that helps WiMax to enhance its coverage area. Another part of this research is to explore P2MP communications and how we can stable the data rate in diverse network. This study provides graphs among distance, BEP and capacity. Finally we can say that WiMax future lies in millimeter wave generating technology called HEMT. We have found that 64QAM modulation technique is good for Pakistani cities and generate less bit error rate than others.

**REFERENCES:**
Arafat, O. and K. Dimyati, (2010) "Performance Parameter of Mobile WiMax: A Study on the Physical Layer of Mobile WiMAX under Different Communication Channels and Modulation Technique" in the proceedings of 2010 Second International